# Performance Evaluation of IEEE 802.11 for UAV-based Wireless Sensor Networks in NS-3

Md. Abubakar Siddik, Md. Rajiul Islam, Md. Mahafujur Rahman, Zannatul Ferdous, Sumonto Sarker, Most. Anju Ara Hasi and Jakia Akter Nitu

*Department of Electronics and Communication Engineering, Hajee Mohammad Danesh Science and Technology University, Dinajpur, Bangladesh*

***ABSTRACT :*** *Unmanned Aerial Vehicle (UAV) has extreme potential to change the future wireless sensor network (WSN). The most efficient, high-performance, urgent and fastest data collection are achieved in WSN through the UAV because UAV provides an energy-efficient short-range communication with good connectivity. However, UAV-based WSN performances is influenced by different system parameters. To investigate this issue, it is necessary to analyses the effects of system parameters on the UAV-based WSN performance. In this paper, we design a NS-3 script for UAV-based WSN according to the hierarchical manner of TCP/IP model. We configure all layers by using NS-3 model objects and set and modify the values used by objects to investigate the effects of system parameters (access mechanism, UAV trajectory pattern, UAV velocity, number of sensors, and sensor traffic generation rate) on throughput, and average delay. The simulation results show that the RTS/CTS access mechanism provides better performance than the basic data access mechanism and the mobility model which has been prescribed shows higher performance than the random mobility model. Moreover, the results indicate that higher velocity of UAV degrades the system performance in terms of throughput and delay. Our design procedure represents a good guideline for new NS-3 users to design and modify script and results greatly benefit the network design and management.*

---

## I. INTRODUCTION

In the era of advanced technology, unmanned aerial vehicle (UAV known as drone) has become a popular term for enhancing communication technologies as well as for boosting different development and manufacture efforts at a rapid rate [1], [2]. It is widely using in the military, environmental monitoring, industrial, agriculture, surveillance and critical remote area monitoring [3] – [6]. Conversely, the wireless sensor network (WSN) has made an increasing attention for the researchers, industries and governments [7]. The WSN has also wide range applications such as real time monitoring, health care, surveillance, environmental monitoring, and agriculture [8] – [10]. At presents, UAV is used in the WSN data collection which increases the WSN performance in addition for the emergency and critical situation UAV-based WSN determined a promising solution for communication networks [11]. Moreover the UAV-based WSN has some unique characteristics compared to VANET and MANET [12] that have created some challenges such as UAV path and velocity planning, nodes density, sensor nodes transmission policy, positioning of UAV and sensor nodes, hidden node problem, UAV altitude, channel modelling, traffic generation rate and security [13], [14]. The performance of UAV-based WSN depends on these challenges. So, it is an urgent needs to evaluate the UAV-based WSN performance in terms of different system parameters.

In this paper, we evaluate the impacts of UAV path and velocity, number of sensor nodes, access mechanism and traffic generation rates. Figure 1 shows a UAV-based WSN where the sensor nodes are distributed on the ground like a grid position. The UAV is visiting this area from the ground-based control station. The connection between the nodes and UAV is only possible when the distance between them is within their signal transmission range. The UAV continuously moving from Position A to B so the UAV coverage range also continuously shift and it covers new nodes from time to time. Due to the movement of the UAV, it is very difficult to make a connection between the UAV and the sensor nodes. Again, when the number of nodes is increased within the UAV coverage range there needs an effective congestion control mechanism. Because congestion of multiple nodes during data transmission degrades the network performance [15]. To mitigate this difficulty IEEE 802.11a standard is used. IEEE 802.11a is a CSMA/CA for MAC protocol where the frequency is 5 GHz, an orthogonal frequency division multiplexing (OFDM) for physical interface and a distributed coordination function (DCF) MAC function [16]. The DCF includes two access mechanisms one is two-way handshaking access mechanism also known as a basic access mechanism and another is four-way handshaking mechanism known as Request to Send, Clear to Send (RTS/CTS) access mechanism. A four-way handshaking access mechanism avoids the hidden terminal problem and enhances the network performance [17]. In the basic access mechanism, the nodes turn into active mode and take a random backoff number after receiving a beacon signal from UAV. When one node backoff is reduced to zero it waits for distributed interframe space (DIFS) time and then senses the channel. If the channel is still clear the node sends the data to the UAV. The UAV is replied by an ACK to the sending node after the short inter-frame space (SIFS) time. Others nodes then again take the random backoff and repeat this data transmission technique [15]. The basic access mechanism is not capable to reduce the hidden nodes problem. Hidden nodes problem occurs when the multiple nodes can't hear each other within the UAV coverage range [18]. The RTS/CTS mechanism effectively solves the hidden nodes problem by sending an RTS frame to the UAV. When the UAV broadcasts a CTS frame to the nodes. All of the nodes receive the CTS frame and check whether its own frame or not. Then only the successful node sends his data. Again, appropriate route selection of UAV is an important challenge in UAV-based WSN. To evaluate the UAV-based WSN performance in terms of UAV route we are used Gauss Markov and random direction 2D mobility model. Where, the Gauss Markov mobility model calculates the velocity and direction of UAV by the time stem and only one tuning parameter called $\alpha$ [19], [20]. When $\alpha = 1$, the new speed and direction of UAV is identical. Conversely, the random direction 2D mobility model provides random UAV direction of UAV with a constant velocity. Here, when the UAV reaches at the boundary, it pauses and selects a new direction. In Figure 2 it's shown that a random direction two dimensional mobility model. In this figure, the drone moves in a forward direction with a constant speed until reach the boundary. After reach the boundary the drone take a new direction pattern. The details design of the mobility model has given in to the system design section. We conduct the network simulator-3for the system design which is a power simulation tool for real time system [21].

The performance analyses of UAV-based WSN are attractive interests to the researchers which are described as follows. Sun et al. [22] have analyzed the slotted CSMA/CA performance in the UAV-based WSN. They have considered the beacon frame, velocity of UAV, nodes density, and the number of packets for evaluating the network throughput and the remaining number of packets of the nodes. Again, Shuhang Liu et al. [23] have analyzed the UAV-assisted WSN data collection where the entire area is divided into multiple small cells and used a single UAV or multiple UAVs to cover all the cells. They have considered the number of cells and evaluate the single and multiple UAVs impacts on the node capacity. S. Sotheara et al. [24] have proposed a priority-based contention window adjustment scheme protocol based on two optimized frames called Priority-based Optimized Frame (POFS) and Circularly Optimized Frame (COFS). Here, data will be transmitted at first from the rear position higher priority nodes to the UAV. However, this protocol reduces the packet loss but the frequent calculation of the contention window drains energy and time of the network. Moreover, they did not consider the velocity and trajectory of the UAV. Again, S. Say et al. [25] have introduced a partnership-based data forwarding model where nearly positioned nodes make a partnership, and if one node fails to send data then it sends its remainder data to the partner and then the partner will send his data. However, these models significantly reduce the packet loss and huge delay which improve the performance of the network but in the dense network the performance degrades. In [26], has designed an easy MAC protocol for UAV-based WSN data collection. This protocol is based on some frames transmission before the actual data transmission. Although, this protocol ensures a fair chance for every node and the success rate is 100 nevertheless. This protocol has considered the hidden node problem but did not define the UAV route appropriately. Obviously, [24] − [27] proposed channel access mechanisms enhances the network performance. But they did not show the impacts of velocity and trajectory of UAV and traffic generation rates.

In this paper, we develop a NS-3 script for a UAV-based WLAN in NS-3 which consists of an access point (AP) node acts as a UAV and a number of STA nodes act as sensors forming a one-hop star topology. We assume that each node of the network uses IEEE 802.11a standard specifications as MAC and PHY layer, TCP/IP protocol stack as network and transport layer and on off application and packet sink application as application layer. Moreover, we set and modify the attributes of different layer objects to investigate the effects of system parameters (access mechanism, number of sensors, UAV trajectory pattern, UAV velocity, and sensor traffic generation rate) on throughput, and average delay. Most of the previous works focus on the investigation of the effects of system parameters on performance but, to the best of our knowledge, none of the research work has described the design procedure of NS-3 script in accordance with the instructions of TCP/IP model to their manuscript.

The rest of the paper is organized as follows: The configuration of UAV-based WSN based on TCP/IP network model in NS-3 is described in details in Section II. Section III derives the expression of throughput and average delay based on flow monitor attributes of NS-3. The effects of system parameters on throughput and average delay are presented in Section IV. Finally, Section V concludes the work and gives the future work outlines.

## II. SYSTEM CONFIGURATION IN NS-3

In this section, we outlet the configuration of UAV-based wireless sensor networks based on TCP/IP network model in NS-3 and the measurement process of performance metrics using C++ language. Moreover, we present the hierarchical design procedure of WLAN in NS-3 and also highlight how the attributes or parameters of different NS-3 models or classes can modify to design a new configuration of a network. The NS-3 is a well-organized and maintained, flexible and simple architecture, easy and accessible documentation, fully open source, and widely used in academic and research as a network simulation tool and it has high accuracy and speedy execution capability to run NS-3 scripts. NS-3 provides different types of helper API, container API and core API to design a complete communication system. The container API performs a number of identical actions to groups of objects and the core API performs a particular task for a NS-3 model. The helper API is used to write and read NS-3 script easier. We first include the necessary namespaces and header files at top of the NS-3 script. The rest of the part of the NS-3 script is described as follows:

### A. Physical Topology Configuration

In this work, we consider a UAV-based WSN consists of an UAV that acts as an AP node and a number of identical STA nodes in a star topology to evaluate the performance in terms of throughput and delay. In order to create an AP node and a number of STA nodes, we take two objects of *NodeContainer* class, named as *apNode*, *staNode,* and use *create ()* function which takes the number of AP nodes and number of STA nodes as a parameter. To assign positions and to set mobility pattern to the AP and STA nodes, we take two objects of *MobilityHelper* class, named as *apMobility*, *staMobility,* and use *SetPositionAllocator () and SetMobilityModel ()* functions. The NS-3 provides different position allocators and mobility models to describe the network topology properly. Each position allocators and mobility models can set position and mobility pattern to the nodes according to a set of attributes. Finally, the defined properties of position allocator and mobility model are assigned to AP and STA nodes by using *Install ()* function which takes *apNode* and *staNode* objects as a parameters, respectively. In this work, we consider *ConstantPositionMobilityModel* class as a mobility model for STA and either *GaussMarkovMobilityModel* or *RandomDirection2dMobilityModel* are used as a mobility model for UAV.

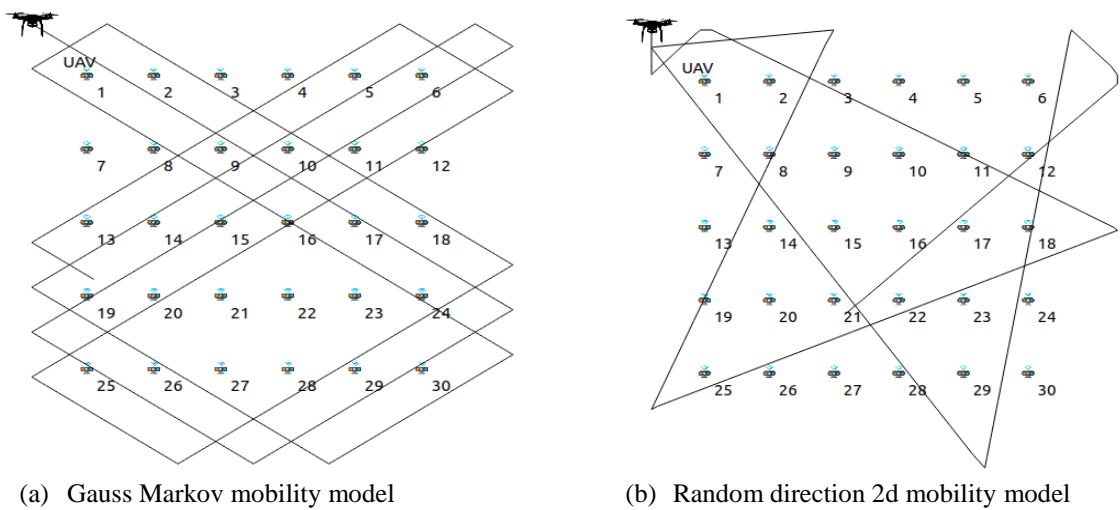

(a) Gauss Markov mobility model　　　　　　(b) Random direction 2d mobility model

Figure 1: PHY topology of UAV-based WSN generated in NS-3

### B. Network Device Configuration

The *NetDevice* class of NS-3 provides objects as network interface card of actual computer for Ethernet, WiFi, Bluetooth, etc. In this paper, we consider *WifiNetDevice* class to design IEEE 802.11-based WLAN. The *WifiHelper* class is used to create *WifiNetDevice* objects for each node which is the interface between network layers to data link layer. The common properties of *WifiNetDevice* like remote station manager and PHY standard for both AP and STA nodes are assigned by using *SetRemoteStationManager ()* and *SetStandard ()* functions through *WifiHelper* class object named *wifi*. To configure differentiable properties of *WifiNetDevice* for AP and STA nodes, we create two objects of *NetDeviceContainer* class, named as *apNetDevice* and *staNetDevice*. Finally, *WifiNetDevice* for AP and STA nodes is configured by using *wifi* object and *Install ()* function which takes *WifiPhyHelper* object, *WifiMacHelper* object and *NodeContainer* object as a parameters. In this work, to show the effects of access mechanism, we modify the *RtsCtsThreshold* attribute value using *SetRemoteStationManager ()* function.

### C. PHY Layer Configuration

In order to configure *WifiPhy* model as the PHY layer, we take an object of Yans*WifiPhyHelper* class, named as *wifiPhy*, because both type of nodes (AP and STA) use same PHY layer. Moreover, we also assign error model to the PHY layer object by using *SetErrorRateModel ()* function. Then we take an object of Yans*WifiChannelHelper* class, named as *wifiChannel*, to configure the channel for both AP and STA nodes. The *AddPropagationLoss ()* functions is used to set propagation loss model to the channel object and finally, this channel is assigned to PHY layer through *SetChannel ()* function.

### D. MAC Layer Configuration

To configure MAC layer model for the AP and STA nodes, we take into account two objects of *WifiMacHelper* class, named as *apWifiMac*, *staWifiMac*, because AP and STA nodes need different configurations to communicate with each other. The NS-3 offers three types of MAC model like *ApWifiMac*, *StaWifiMac* and *AdhocWifiMac*. To set *ApWifiMac* model as the MAC layer of AP node, we use *SetType ()* function and assign different MAC parameters values as attributes to change the configuration as we needed. Similarly, configuration of MAC layer of STA nodes is performed by using *StaWifiMac class and SetType ()* function.

### E. Network and Transport Layer Configuration

To configure network and transport layer of AP and STA nodes, we consider an object of *InternetStackHelper* class, named as internetStack, because both AP and STA nodes use same network and transport layer protocols. We set all default configurations of network and transport layer to *apNode*, *staNode*objects by using *Install ()* function. *InternetStackHelper* class aggregates IP/TCP/UDP functionality viz. *ArpL3Protocol*, *Ipv4L3Protocol*, *Ipv6L3Protocol*, *Icmpv4L4Protocol*, *Icmpv6L4Protocol*, *UdpL4Protocol*, *TrafficControlLayer*, *PacketSocketFactory*, *Ipv4* routing, *Ipv6* routing, by default, to each node. Finally, we finish the network layer configuration process after the end of the IP address set and assign procedure. In order to assign IP address to each network device of AP and STA nodes (apNetDevice and *staNetDevice*), we create an object of *Ipv4AddressHelper* class, named as *ipAddr*. We use *SetBase ()* and *Assign ()* functions to set IP address for AP and STA nodes and to assign this IP address through two objects of *Ipv4InterfaceContainer* class, named as *apNodeInterface* and *staNodeInterface*.

### F. Application Layer Configuration

To configure application layer, we need application container where the protocols of application layer are installed. We take two objects of *ApplicationContainer* class, named as *apApplication* and *staApplication*. NS-3 offers different types of applications by defining different application class. In this work, we use *OnOffApplication* class as the application of STA nodes and *PacketSink* class as the application of AP node. Socket address (IP address of network layer protocol and port number of transport layer protocol) and transport layer protocol of the node are needed to configure an application. We take two objects of *InetSocketAddress* class, named as *apAddress* and *staAddress*, which contain the socket address of AP and STA nodes. Then, we consider an object of *PacketSinkHelper* class, named as *apSink*, and after that, *SetAttribute ()* function is used to define the attributes of the packet sink application. Finally, *Install ()* function is used to install the *PacketSink* application into the AP node. Similarly, we take an object of *OnOffHelper* class, named as *staOnOff*, and *SetAttribute ()* and *Install ()* functions are used to configure the *OnOffApplication* to each STA nodes. Finally, we initialize the start and stop time of the applications of AP and STA nodes by using *Start ()* and *Stop ()* functions of *ApplicationContainer* class. The traffic generation rate of a node can modified by changing the value of *DataRate* attribute. In this work, to show the effects of traffic generation rate, we use three values of *DaraRate* attributes: 0.5 Mbps, 3Mbps and 6 Mbps.

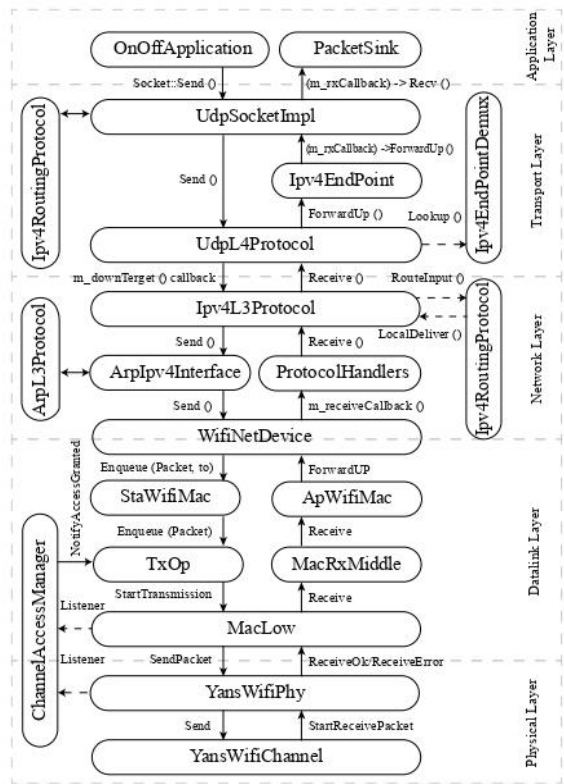

Figure 2: Communication architecture of UAV-based WSN in NS-3

### G. Simulation Setup

The *Simulation* class of NS-3 executes the simulation events and control the virtual time. In this work, we use three static functions of *Simulation* class to execute our designed NS-3 script for *simulationTime* time. The *Run ()* function runs the simulation and will be continued until at least one of the three events occurs: (a) no events are present anymore, (b) user called *Stop ()* function, and (c) user called *Stop ()* function with a stop time and the expiration time of the next event to be processed is greater than or equal to the stop time. The *Destroy ()* function is typically called at the end of a simulation to avoid false-positive reports by a leak checker. After this method has been called, it is actually possible to restart a new simulation. The *Stop ()* function tells the *Simulator* class the calling event should be the last one executed. It takes the *simulationTime* as a function parameter.

### H. Animation Setup

The NS-3 has two ways to provide animation, namely using the *PyViz* method or the *NetAnim* method. In this paper, we use *NetAnim* to display the topology of the network and animate the packet flow between nodes. The *NetAnim* also provides useful features such as tables to display meta-data of packets. To trace the file generated during simulation, we use an object of *AnimationInterface* class, named as *animation*. *AnimationInterface* class traces the statistics for each flow and exports in XML format.

### I. Event Monitoring and Data Collection

The NS-3 supports many event monitoring model like ASCII trace, PCAP and flow monitor. In this work, we use flow monitor model to measure the performance of network protocols. The flow monitor is a flexible event monitoring model and it uses probes to track the packetsat IP level exchanged by the nodes and the packets are divided according to the flow they belong to, where each flow is defined as the packets with same (protocol, source IP address, source port number, destination IP address, destination port number) tuple. Flow monitor collects the statistics for each flow and exports in XML format. To enable the flow monitor, we take an object of *FlowMonitorHelper* class, named as monitor and use *InstallAll ()* function. The data produced during simulation is traced by *FlowMonitor* class and flow monitor model stores this data in a map according to flow id using variables: *timeFirstTxPacket, timeLastTxPacket, timeFirstRxPacket, timeLastRxPacket, delaySum, jitterSum, txBytes, rxBytes, txPackets, rxPackets, timesForwarded, bytesDropped, packetsDropped*, etc. We manipulate this data and measure the throughput and average delay according to the definition defined in Section IV. After calculating performance metrics, we create a CSV file and write the value of performance metrics to the file using *ofstream* class of C++.

### III. PERFORMANCE METRICS

In this section, we outline how the principal performance metrics, viz. throughput, and average delay, are determined from flow monitor attributes.

**A. Throughput**

The throughput of a STA node or a flow is defined as the number of bits of the STA node successfully received by AP node in unit time. Throughput is measured in bits per second (bps). The throughput of the network is the summation of all indivudual STA node throughput. Therefore, if the number of STA nodes in the network is $n$, the throughput of the network is expressed as:

$$\text{Throughput} = \sum_{i=1}^{n} \text{Throughput of flow i} = \sum_{i=1}^{n} \frac{\text{Received bytes of flow i}}{\text{Simulation time}} = \sum_{i=1}^{n} \frac{i->second.rxBytes \times 8}{simulationTime} \quad (1)$$

**B. Average Delay**

The average delay is defined as the ratio of the sum of all end-to-end delays for all received packets and total number of received packets. The average delay is measured in seconds (sec). Therefore, if the number of STA nodes in the network is $n$, the average delay is expressed as:

$$\text{Average delay} = \sum_{i=1}^{n} \frac{\text{Sum of packet delay of flow i}}{\text{Received packets of flow i}} = \sum_{i=1}^{n} \frac{i->second.delaySum.GetSeconds}{i->second.rxPackets} \quad (2)$$

### IV. SIMULATION RESULTS

In this section, we evaluate the effects system parameters viz. number of nodes, access mechanism, node velocity, mobility model and traffic generation rate, on performance metrics of one-hope star topology structure in a UAV-based WSN under the IEEE 802.11a standard. The simulation experiments were conducted using NS-3 (version 3.30). In the simulation experiments, we considered three network scenarios (indexed by 1, 2 and 3) that are used to investigate the effects of five system parameters. The simulation experiment consists of an AP node and a number of STA nodes that adopt all default configurations defined in NS-3 under IEEE 802.11a standard except few attributes which are given in Table 1.

**A. Impact of Number of Sensors and Access Mechanism**

In order to investigate the impact of number of sensors and access mechanism on the throughput and average delay, we consider a network scenario that contains up to 60 sensors where traffic generation rate of each node is 6 Mbps and a UAV that moves according to Gauss Markov mobility model with 50 m/s velocity. The network operates with either basic or RTS/CTS access mechanism while other parameters are same as general scenario. The impact of sensor density varies from 0 to 60 and access mechanism (either basic or RTS/CTS) on the throughput and average delay are illustrated in Figure 3.The result in Figure 3 (a) shows that the throughput exponentially increases up to 20 sensors for both access mechanism. This is because the collision probability increases slowly compared to transmission probability with the number of sensors increasing within a certain range. It is observed that the throughput gradually decreases as the number of sensors increases after 20 sensors because with the number of sensors further increasing, the collision probability increases dramatically. The result also shows that the RTS/CTS access mechanism achieves better throughput compared to basic access mechanism. This is due to the fact that the increasing rate of collision probability of the basic access mechanism is more than the RTS/CTS access mechanism for dense network. The Figure 3 (b) shows that the average delay exponentially increases as the number of sensors increases for both access mechanism. This is because the collision probability increases with the increase of number of sensors and thus increase of access delay. The result also shows that the basic access mechanism suffer from higher average delay compared to the RTS/CTS access mechanism. This is due to the fact that the increasing rate of collision probability of the basic access mechanism is more than the RTS/CTS access mechanism for dense network.

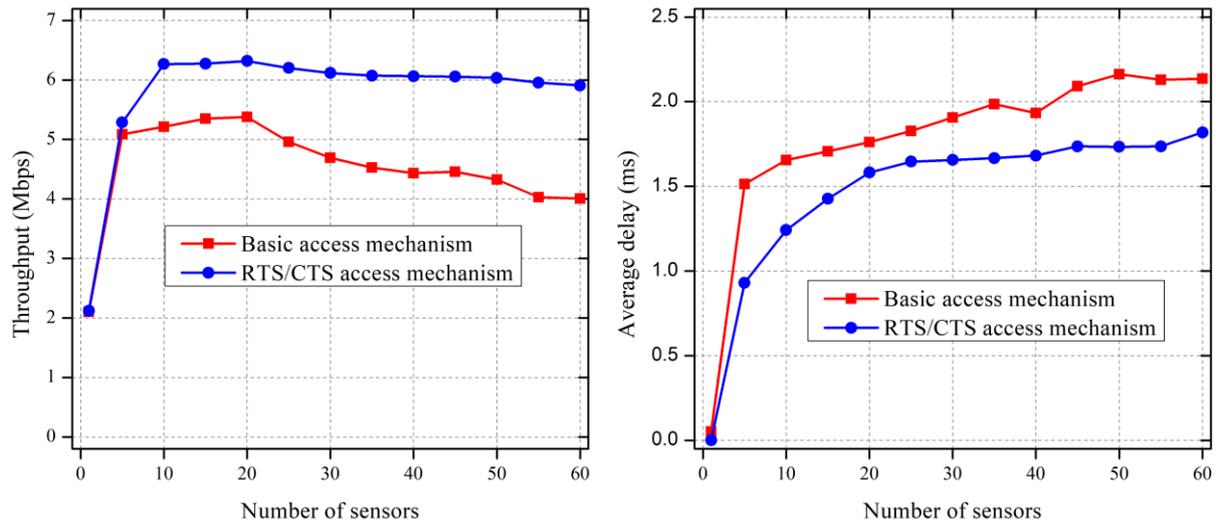

**Fig.3. Impact of number of sensors and access mechanism on (a) throughput and (b) average delay**

Table1. Simulation attributes

| Configuration | Attributes | Values |
|---|---|---|
| General | Mobility model for STA nodes | *ConstantPositionMobilityModel* |
| | Remote station manager | *ConstantRateWifiManager* |
| | Data mode and control mode | *OfdmRate12Mbps* |
| | Wifi channel | *YansWifiChannel* |
| | Propagation loss model | *RangePropagationLossModel* |
| | Wifi PHY | *YansWifiPhy* |
| | Error rate model | *YansErrorRateModel* |
| | Wifi MAC for AP node | *ApWifiMac* |
| | Wifi MAC for STA nodes | *StaWifiMac* |
| | Number of AP nodes | 1 |
| | Simulation time | 30 sec |
| | Event monitoring model | *FlowMonitor* |
| | Animation model | *NetAnim* |
| Scenario 1 | RTS threshold | {0, 65535} |
| | Number of STA nodes | [1,60] |
| Scenario 2 | Mobility model for AP node | *GaussMarkovMobilityModel* |
| | | *RandomDirection2dMobilityModel* |
| | Number of STA nodes | [1,60] |
| Scenario 3 | Traffic rate | {0.5, 3.0, 6.0} Mbps |
| | Velocity | [1,70] m/s |

**B. Impact of Number of Sensors and UAV Trajectory Pattern**

In order to investigate the impact of number of sensors and UAV trajectory pattern on the throughput and average delay, we consider a network scenario that contains up to 60 sensors where traffic generation rate of each node is 6 Mbps and a UAV that moves according to either Gauss Markov mobility model or random direction 2d mobility model with 50 m/s velocity. The network operates with RTS/CTS access mechanism while other parameters are same as general scenario. The impact of number of sensors varies from 0 to 60 and mobility model (either Gauss Markov or random direction 2d) on the throughput and average delay are illustrated in Figure 4. The result in Figure 4 (a) shows that the throughput exponentially increases up to 20 sensors for both Gauss Markov mobility model and random direction 2d mobility model. This is because the collision probability increases slowly compared to transmission probability with the number of sensors increasing within a certain range. It is observed that the throughput gradually decreases as the number of sensors increases after 20 sensors because with the number of sensors further increasing, the collision probability increases dramatically. The result also shows that the Gauss Markov mobility model provides a slightly better result compare to the random direction 2D mobility model in the dense network. This is because the Gauss Markov mobility model provides users with defining appropriate trajectory patterns. So every node within the UAV coverage area gets an equal chance to send their packets. In contrast, the random direction 2D mobility model does not provide an appropriate user define route pattern. It takes a random trajectory.As a result, some nodes get more chances to send their packets whether other nodes deprive of sending their packets. The Figure 4 (b) shows that the average delay exponentially increases as the number of sensors increases for both Gauss Markov mobility model and random direction 2d mobility model. This is because the collision probability increases with the increase of number of sensors and thus increase of access delay. The result also shows that the random mobilitymodel shows slightly higher average delay than the Gauss Markov mobility model due to the exact route pattern of the Gauss Markov mobility model.

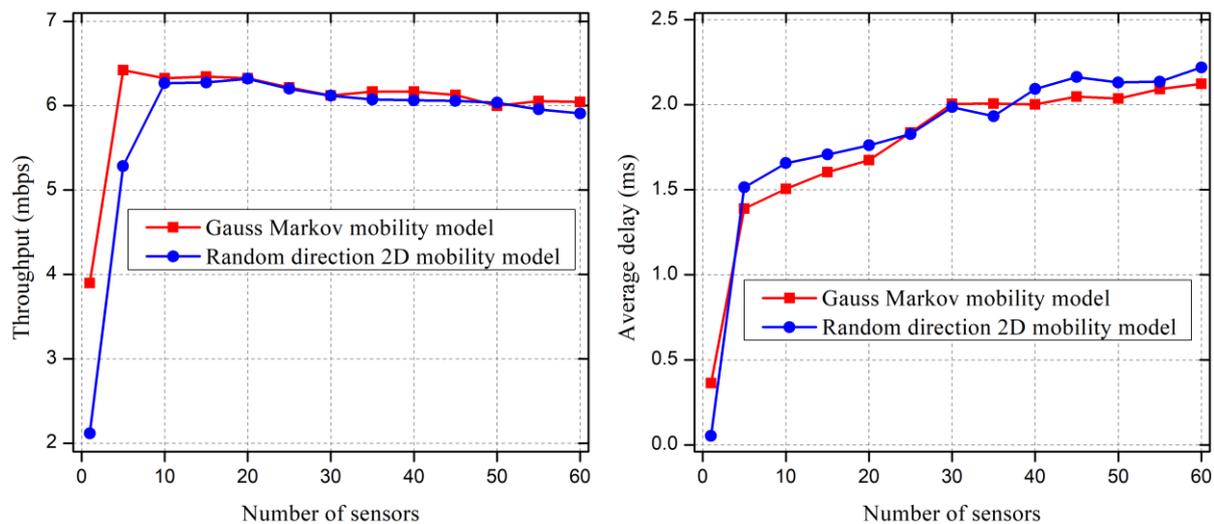

**Fig.4. Impact of number of sensors and UAV trajectory pattern on (a) throughput and (b) average delay**

**C. Impact of UAV Velocity and Sensor Traffic Generation Rate**

In order to investigate the impact of UAV velocity and sensor traffic generation rate on the throughput and average delay, we consider a network scenario that contains 25 sensors and a UAV that moves according to Gauss Markov mobility model. The network operates with RTS/CTS access mechanism while other parameters are same as general scenario. The impact of UAV velocity varies from 0 m/s to 70 m/s and sensor traffic generation rate (either 0.5 Mbps, 3 Mbps or 6 Mbps) on the throughput and average delay are illustrated in Figure 5. The result in Figure 5 (a) shows that the throughput of nodes increases with increasing the UAV velocity up to 10 m/s for 3 and 6 Mbps traffic generation rateand 30 m/s for 0.5 Mbps then it starts to decrease slowly This is because, when the UAV speed is too high the residencetime is less as consequence nodes are failing to send all packets.

Conversely, when the UAV velocity is too low it does not collect all nodes packets within the limited simulation time as a result throughput is also lower at the low velocity. Besides,the throughput of nodes is slightly higher for higher traffic generation rate. The Figure 5 (b) shows that the average delay of nodes becomes stable as the velocity of the UAV increases. This is because with the velocity of UAV increasing the distance between UAV and nodes within the coverage area remains constant. Moreover, nodes with higher traffic generation rates suffer from more delays than lower traffic generation rate nodes. Due to increases in the traffic generation rate traffic load of the nodes increases but the service rate remain unchanged as a result more queuing delay raisedin the node's queue consequently average delay of the nodes increases.

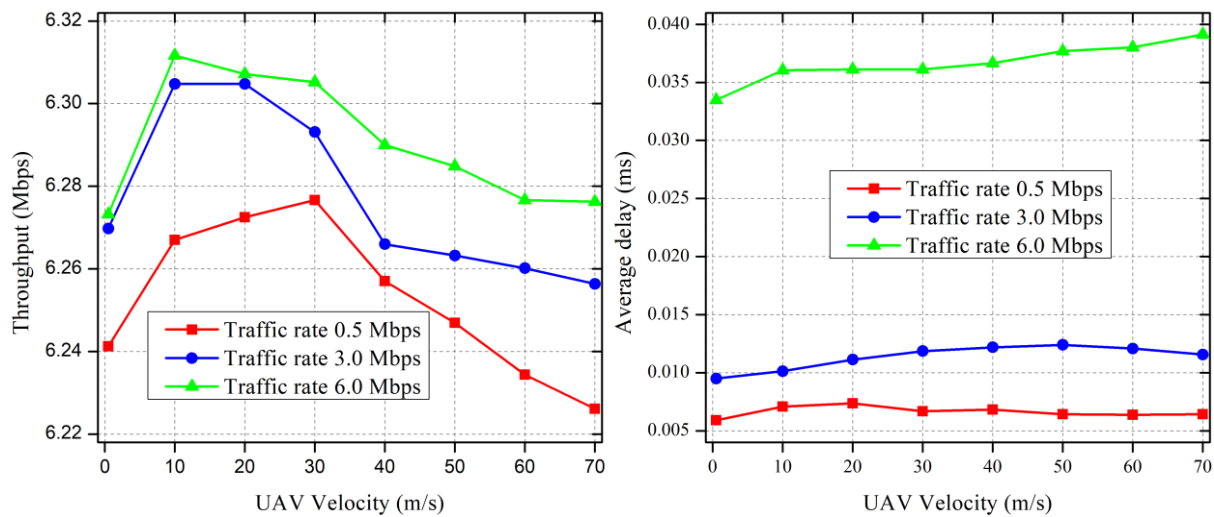

**Fig. 5. Impact of UAV velocity and sensor traffic generation rate on (a) throughput and (b) average delay**

## V. CONCLUSION

In this paper, we designed a NS-3 script to investigate the effects of system parameters on throughput and average delay for UAV-based WSN in NS-3 and organized it according to a TCP/IP model. The determination of throughput and average delay in context of NS-3 attributes is also discussed in this work. Unlike the previous related works, this hierarchical design procedure of UAV-based WSN in NS-3 provides a complete guideline for new user of NS-3. Moreover, we investigated the effects of system parameters (access mechanism, number of sensors, UAV trajectory pattern, UAV velocity, and sensor traffic generation rate) on throughput, and average delay. The simulation results show that the combination of RTS/CTS data collection technique and effective Gauss Markov mobility model enhance the performance of UAV-based wireless sensor network. We believe that the simulation results would assist the protocol developers to design effective and efficient protocols and to select optimal value of different system parameters to enhance the performance. In future, we will explore the effects of different system parameters on both DCF and EDCA mechanisms in UAV-based WSN.